\documentclass{iaus_ltr}
\usepackage{graphicx}

\title[SEDs of mid-IR selected quasars] 
{Spectral energy distributions of quasars selected in the mid-infrared}

\author[Lacy et al.]   
{M.\ Lacy$^1$, A.\ Sajina$^2$, A.O.\ Petric$^3$, S.E.\ Ridgway$^4$, D.M.\ Nielsen$^5$, T. Urrutia$^6$, D.\ Farrah$^7$, E.L.\ Gates$^8$}

\affiliation{$^1$National Radio Astronomy Observatory, \\ 520 Edgemont Road, Charlottesville, VA22903, USA \\ email: {\tt mlacy@nrao.edu} \\[\affilskip]
$^2$Dept. of Physics and Astronomy, Tufts University\\ Medford, MA02155 \\[\affilskip]
$^3$Department of Astronomy, California Institute of Technology\\ 
1200 E. California Boulevard, Pasadena, CA91125, USA\\[\affilskip]
$^4$National Optical Astronomy Observatory\\ 950 N. Cherry Avenue, Tucson, AZ85719, USA\\[\affilskip]
$^5$Dept. of Astronomy, University of Wisconsin,\\Madison, WI53706 \\[\affilskip]
$^6$Liebnitz Institut f\"{u}r Astrophysik Astrophysics Potsdam\\
An der Sternwarte 16, 14482, Potsdam, Germany \\[\affilskip]
$^7$Department of Physics and Astronomy, University of Sussex\\ Falmer, Brighton BN1 9QH, UK\\[\affilskip]
$^8$UCO/Lick Observatory, University of California\\1156 High Street, 
Santa Cruz, CA95064 \\}

\pubyear{2011} 
\volume{284}  
\pagerange{1--12}
\jname{The Spectral Energy Distribution of Galaxies}
\editors{R.J. Tuffs \&  C.C.Popescu, eds.}

\begin{document}

\maketitle

\begin{abstract}
We present preliminary results on fitting of SEDs to 142 $z>1$ quasars selected
in the mid-infrared. Our quasar selection finds objects ranging 
in extinction from highly
obscured, type-2 quasars, through more 
lightly reddened type-1 quasars and normal
type-1s. We find a weak tendency for the objects with the highest far-infrared
emission to be obscured quasars, but no bulk systematic offset between the
far-infrared properties of dusty and normal quasars as might be expected in 
the most naive evolutionary schemes. The hosts of the type-2 quasars have
stellar masses comparable to those of radio galaxies at similar redshifts.
Many of the type-1s, and possibly a
one of the type-2s require a very hot dust component in addition to the normal
torus emission.

\end{abstract}

\firstsection 
\section{Introduction}

Over the past three years we have been conducting an extensive campaign of 
follow-up spectroscopy of dust obscured quasars, using the selection technique
described in Lacy et al. (2004; 2007). We now have over 700 spectra of 
AGN and quasars selected in this manner, forming the {\em Spitzer} 
Mid-InfraRed Quasar Survey (SMIRQS). SED fitting has been a key activity
carried out in parallel with the spectroscopic follow-up. It allows us to 
separate the stellar, AGN and starburst components of the quasar and its host, it enables a consistency check on the spectroscopic redshift
(which may be of low signal-to-noise in the case of highly-obscured objects), 
it helps with classification of $z>1$ objects between type-2 and reddened 
type-2 that lack rest-frame optical (i.e.\ near-infrared) spectra, and it 
allows us to test theories for the origin and nature of dusty quasars and 
their various SED components.

The two 
most popular explanations for the existance of dust-reddened quasars are
orientation, where a dusty nuclear torus provides a large amount ($A_V>10$) of
extinction to the quasar (Antonucci 1993), 
and evolution, where a quasar begins its life 
highly obscured (again $A_V>>10$) following a galaxy merger and ensuing 
starburst, 
but then gradually expels its dusty envelope and 
eventually shines as an unobscured quasar (Sanders et al. 1988; Hopkins et al.\ 2008). The relative roles of orientation and evolution have been a topic of 
some debate, one simple, if somewhat naive, 
test is whether the obscured quasars show significantly
more signs of ongoing star formation than their unobscured counterparts, 
consistent with them being at an earlier stage in the evolutionary sequence.

We have also used our SED fits to study the very hot dust component in the
near-infrared. The strength of this recently-recognized 
component appears to correlate with the bolometric luminosity of the quasar.
Mor \& Trakhtenbrot (2011) use WISE data 
to demonstrate that, because this 
dust is above the sublimation temperature of silicate grains, it must arise 
in pure graphite grains in clouds beyond the dust-free broad-line region, 
but inside the torus.

In this paper we present the results of a preliminary analysis of the 
142 $z>1$ objects in our mid-infrared selected quasar sample (two example
fits are shown in Figure 1). Full results will
be presented in Farrah et al.\ (in preparation) and Petric et al.\ 
(in preparation).

\section{The very hot component}

We confirm the need for a very hot dust component in the near-infrared
spectra of many of our type-1 objects, and
also marginal evidence in one of our type-2s. Extinction is probably 
responsible for the lack of these in most of our dusty type-1 and
type-2 quasars. The apparent need for a very hot component in one
of our type-2s (SW105213.39+571605.0; $z=1.242$) is intriguing, and may help 
to constrain the geometry of the emission from this component.

\section{Stellar masses}

The objects classified as type-2 have had their stellar masses estimated by 
fitting stellar population models from Maraston (2005). We fit models that 
are very 
simple single or dual stellar populations (see Lacy et al.\ 2011 for details). 
The results (Figure 2, left)
show that the mean stellar masses are around $2\times 10^{11}M_{\odot}$, but 
with a wide range, $10^{10-12.4}M_{\odot}$. 
There is a loose correlation with quasar
luminosity. These stellar masses are similar to those seen in radio galaxies
(de Breuck et al. 2010) at these redshifts, somewhat 
surprising given that Kukula et al.\ (2001) claim a significant difference
in the host galaxy luminosities of radio-loud and radio quiet quasars at these
redshifts.

\begin{figure}
\begin{picture}(200,400)
\put(0,20){\includegraphics[width=4.0in]{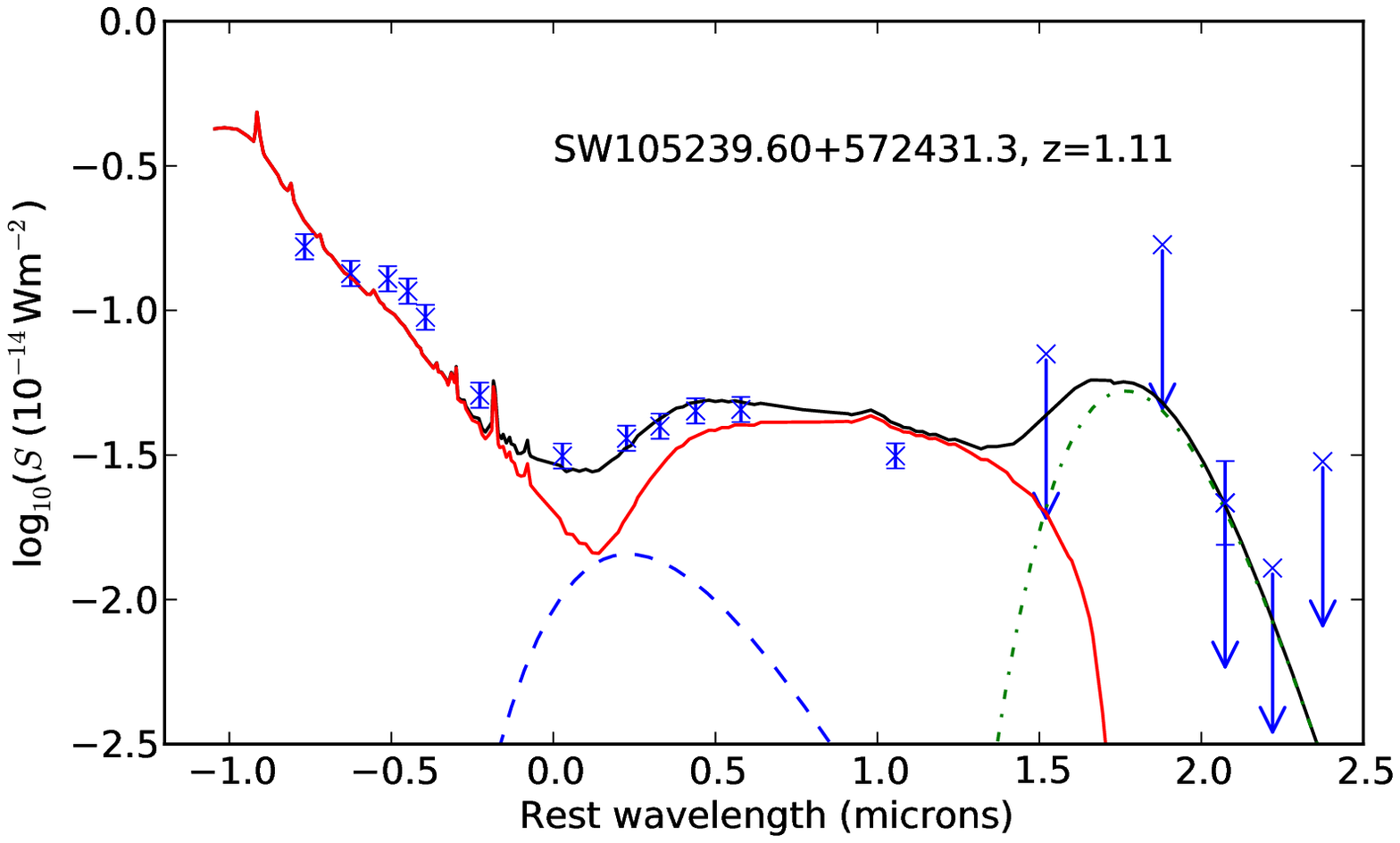}}
\put(0,-180){\includegraphics[width=4.0in]{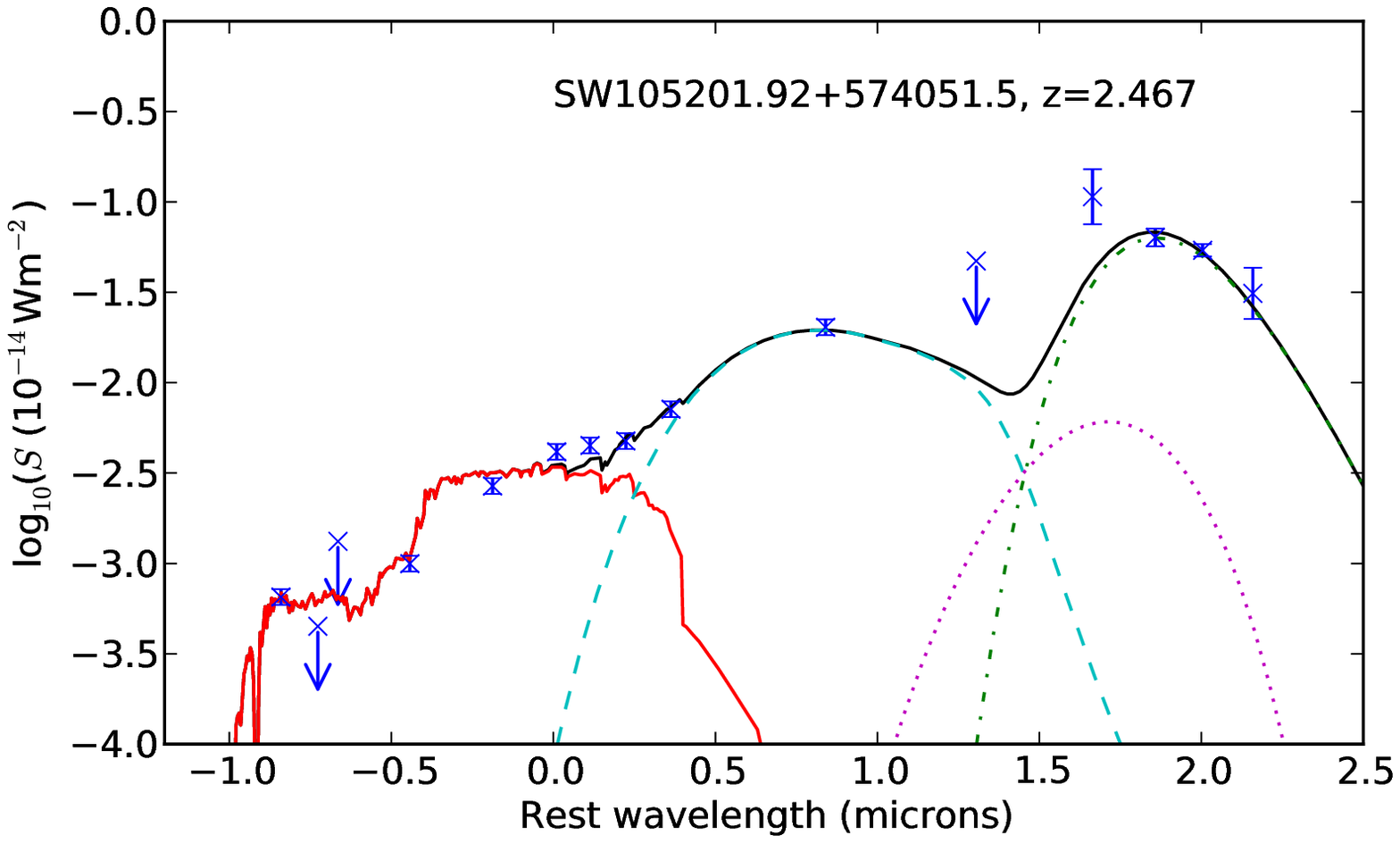}}
\end{picture}
 \caption{Example SED fits: {\em Top} A type-1 quasar SED fit, split into components (from left to right), an optical/IR quasar template (solid), a hot dust bump (dashed line) and a far-infrared modified black body (dot-dashed line). {\em Bottom } a type-2 quasar fit, components (from left to right), stellar light (solid), AGN-heated dust (dashed line), a warm dust component (dotted line) and a far-infrared modified black body (dot-dashed line).} 
   \label{fig1}
\end{figure}

\section{The cool dust component}

Over 50\% of our quasars show a cool dust component, visible as detections
in archival {\em Herschel} data and/or {\em Spitzer} data. The detected objects
have far-infrared luminosities characteristic of Ultraluminous Infrared 
Galaxies (ULIRGS). Dust temperatures are seen
to vary widely, from $\sim 30$ to $> 100$K. We suspect many of the warmer
objects may have contributions to their dust heating from AGN 
(see also Schumacher et al. 2012).

The luminosity of the cold component correlates with the AGN luminosity 
(Figure 2, right) with a logarithmic slope well below unity, as 
also seen in low-z, normal quasars e.g.\ Netzer et al.\ (2007). For the most
part, the dusty type-1 and type-2 quasars fall on the same correlation as the 
normal type-1s, however, all the objects well above the correlation are 
dusty or type-2 objects.

\begin{figure}[b]
\begin{picture}(200,200)
\put(0,0){\includegraphics[width=3.0in]{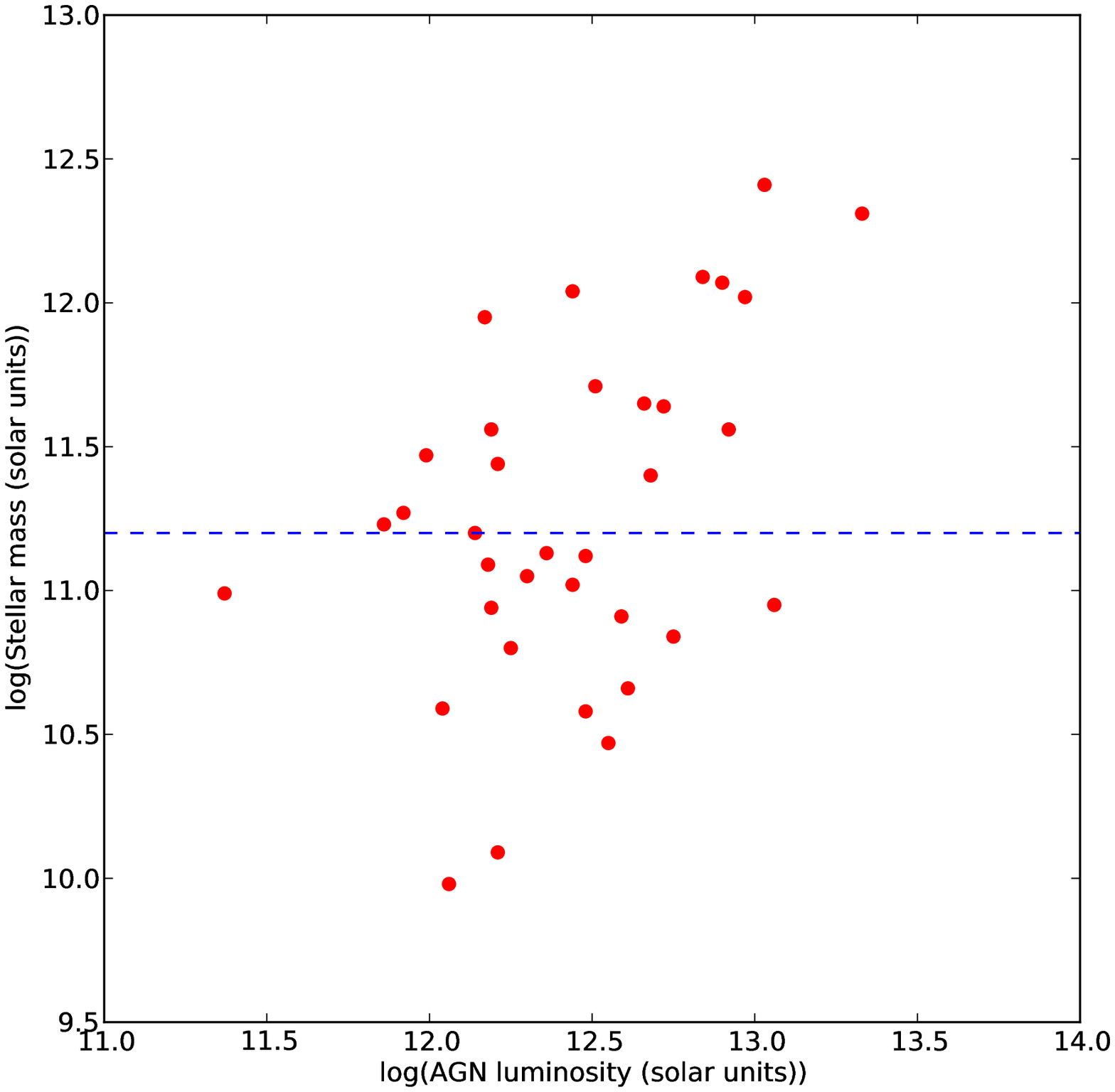}}
\put(200,0){\includegraphics[width=3.0in]{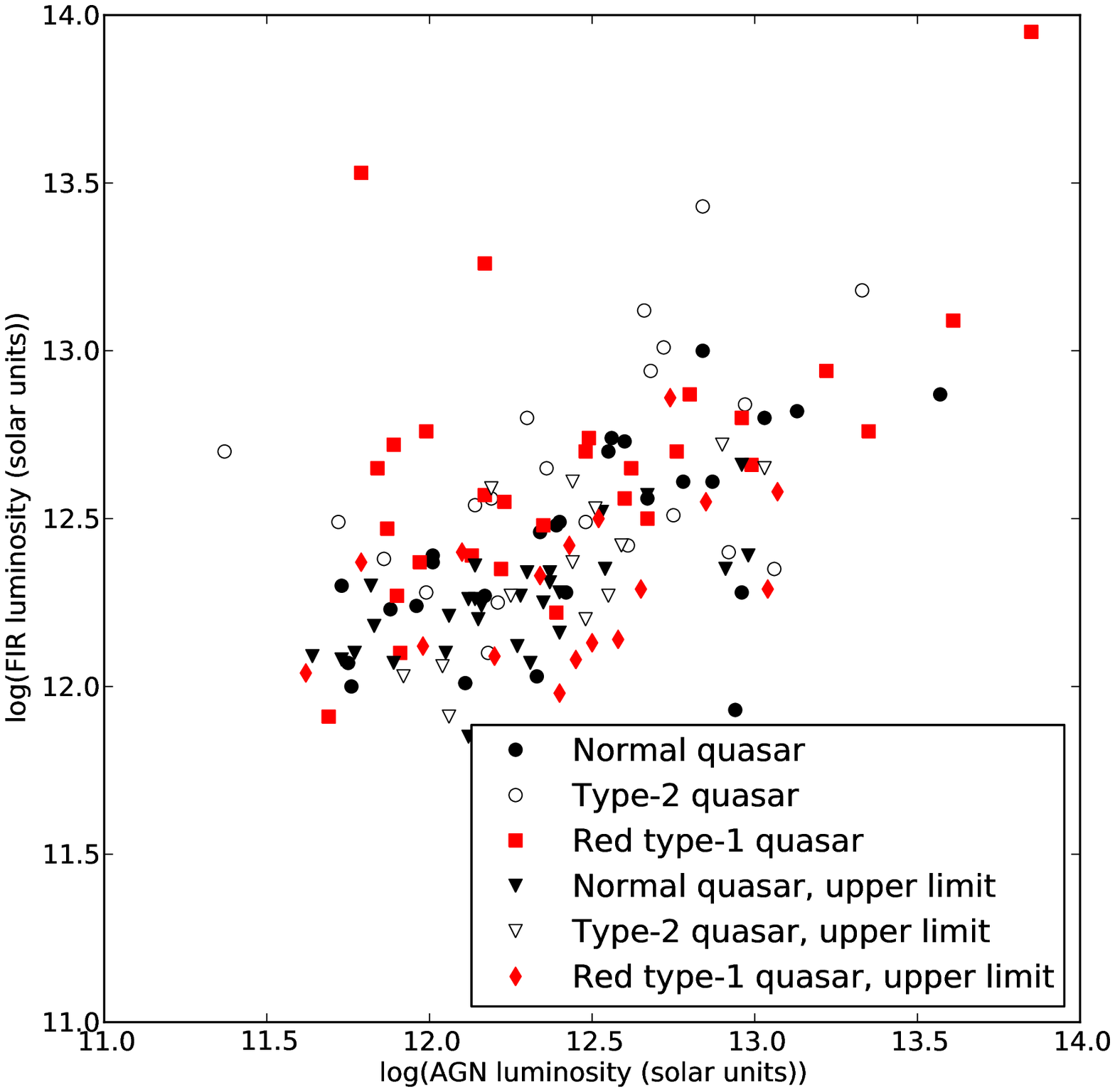}} 
\end{picture}
 \caption{{\em Left} Stellar masses of the $z>1$ type-2 quasars in the sample. The dashed horizontal line shows the mean stellar mass of both this sample and the 
radio galaxy sample of de Breuck et al.\ (2010). {\em Right} Far-infrared luminosity versus AGN luminosity for our quasars.}
   \label{fig1}
\end{figure}

\section{Discussion}

Our results show that more than half of 
mid-infrared selected quasars, regardless of 
optical classification as type-1, reddened type-1 or type-2, are hosted by 
galaxies having the far-infrared
luminosities of ULIRGs, although the amount of far-infrared emission 
contributed by the AGN is unclear in many cases. The most far-infrared 
luminous of these quasars are all obscured to some extent, however, thus
evolutionary schemes, in which dusty quasars are an early stage in the
evolution of quasars are certainly not ruled out by this study.

The stellar masses of the host galaxies of the
type-2 quasars are high, consistent with those
of radio galaxies at the same epoch, suggesting that these quasars 
are relatively mature systems even at $z\sim 2$. Where we have been able 
to estimate CO masses, these seem lower than submm-selected galaxies at 
these redshifts, also consistent with the idea that these are relatively
mature systems (Lacy et al.\ 2011).

\end{document}